\documentclass[a4paper,11pt]{article}
\usepackage{graphicx,amssymb,amsmath,bm,latexsym,color,epsf,cite,verbatim, mathtools}

\usepackage{hyperref}
\usepackage{ulem}

\makeatletter
\@addtoreset{equation}{section}

\makeatother
\newcommand{\be}{\begin{equation}}
\newcommand{\ee}{\end{equation}}
\newcommand{\bea}{\begin{eqnarray}}
\newcommand{\eea}{\end{eqnarray}}
\newcommand{\vs}[1]{\vspace{#1 mm}}

\newcommand{\cD}{{\cal D}}

\newcommand{\bR}{\bar R}

\newcommand{\bnabla}{\bar\nabla}

\newcommand{\Tr}{{\rm Tr}}

\def\eff{\mathbf{f}}
\def\ess{\mathbf{S}}
\def\ti{\mathbf{T}}

\begin{document}

\begin{center}
{\Large\bf Quantum fields and the cosmological constant
\\

}

\vs{8}

{\large
Renata Ferrero\footnote{e-mail address: renata.ferrero@fau.de},
Vincenzo Naso\footnote{e-mail address: vnaso@sissa.it}
and Roberto Percacci\footnote{e-mail address: percacci@sissa.it}$^{,2}$
} \\
\vs{8}
$^1${\em \textit{Institute for Quantum Gravity, Friedrich-Alexander-Universität Erlangen-Nürnberg,}\\\textit{ Staudtstr. 7, 91058 Erlangen, Germany}}

$^2${\em \textit{International School for Advanced Studies, via Bonomea 265, 34136 Trieste, Italy}}

$^3${\em \textit{INFN, Sezione di Trieste, Italy}}

\end{center}
\vs{1}

\setcounter{footnote}{0} 

\begin{abstract}
It has been shown that if one solves self-consistently the semiclassical Einstein
equations in the presence of a quantum scalar field, with a cutoff on the
number of modes, spacetime become flatter when the cutoff increases.
Here we extend the result to include the effect of
fields with spin $0$, $1/2$, $1$ and $2$.
With minor adjustments, the main result persists.
Remarkably, one can have positive curvature even if the
cosmological constant in the bare action is negative.
\end{abstract}

\normalsize

\tableofcontents

\section{Introduction}

The cosmological constant problem consists of various questions \cite{straumann}.
The first and most striking one concerns the effect of vacuum fluctuations
on the curvature of spacetime.
If one thinks of a quantum field as a large number of oscillators,
the vacuum energy is the sum of their ground state energies.
Since a curved manifold looks flat at short distances,
the divergences of a quantum field in curved spacetime are
the same as in flat spacetime.
Based on this observation, it is a generally accepted practice
to calculate the expectation value of the Energy-Momentum (EM) tensor
in a fixed (typically flat) metric
and then to insert the result in the semiclassical Einstein equations
\be
R_{\mu\nu}-\frac12 R g_{\mu\nu}+\Lambda g_{\mu\nu}
=8\pi G \langle T_{\mu\nu}\rangle \ ,
\label{eineq}
\ee
where $\Lambda$ is a cosmological constant.
Since the expectation value of the EM tensor
grows quartically with the ultraviolet cutoff,
even a very low cutoff leads to an unacceptably large curvature.
\footnote{Alternatively, if one tries to cancel the divergences with $\Lambda$,
one is faced with a severe fine tuning issue.}
If quantum field theory is valid up to the Planck scale,
the whole universe should be of Planck size.
This is often presented as the greatest discrepancy between theory and observation
in the history of science.
Clearly, there must be something wrong
on the side of the theory.
\footnote{One simple solution is to say that one out of the infinitely many
degrees of freedom of the gravitational field,
namely the total volume of spacetime, is not dynamical.
If the total volume is fixed, it is natural to further fix the whole volume form
by a gauge choice, leading to unimodular gravity.
While this has the desired effect, it is hard to imagine how to test
unimodular vs. ordinary gravity.
In this paper we assume that the total volume is dynamical
and focus on the calculation of the vacuum energy.}

Becker and Reuter have argued that it is inconsistent to use the EM tensor
computed in one metric as the source of another metric:
the metric has a nontrivial effect on the expectation value
of the EM tensor that cannot be ignored and changes the result drastically.
Instead, one has to calculate the EM tensor self-consistently,
in the metric that solves the semiclassical Einstein equations (\ref{eineq})
{\it in the presence of that EM tensor}
\cite{Becker:2020mjl,Becker:2021pwo}.
This is, in general, a very hard task.
As an illustration of how it could work,
Becker and Reuter have considered a maximally symmetric background,
where the only free parameter of the metric is the radius,
or equivalently the scalar curvature $R$.
The concrete calculation (that we shall review in Section \ref{sec:2})
is done in Euclidean space, i.e. on a sphere.
Since the spectrum is discrete, it is natural to put a cutoff $N$ on
the principal quantum number.
For fixed $N$, one can then compute the curvature self-consistently
and surprisingly it decreases like $1/N^2$.
The result has been extended also to the noncompact, negative curvature case
\cite{Banerjee:2023ztr}.

There are some aspects of this calculation that require
some comments.
The cutoff $N$ is dimensionless, instead of having the standard
dimensions of mass.
Furthermore notice that calling it a cutoff is slightly misleading:
it suggests that it is a number that has to be sent to infinity
in order to recover the continuum.
Then, according to the standard quantum field theoretic interpretation,
quantities computed at fixed $N$ would not be physical:
only renormalized quantities, from which the infinities have been subtracted,
would be physical.
As emphasized by Becker and Reuter, here one must adopt
a different interpretation: the cutoff has to be taken seriously
as a physical quantity, each $N$ defining a different world
that could in principle exist as a physical system.\footnote{See \cite{Freidel:2022ryr} for a similar discussion on the role of the cutoff from a different point of view.}
Also note that the equation of motion has to be solved for finite
cutoff. If one wants to do so, the cutoff can be sent to infinity,
but only after the equation of motion has been solved.

Conceptually, this calculation bears some similarity to the ADM
calculation of the self-energy of a massive particle
\cite{Arnowitt:1960zz}, also reviewed in \cite{Ashtekar:1991hf} and  more recently  revisited in \cite{Woodard:2024zds}.
Normally, if we consider a point particle of mass $m_0$ 
to be the limit $\epsilon\to 0$
of a shell of radius $\epsilon$, the energy of its gravitational field
is $m_0-\tfrac{m_0^2}{2\epsilon}$.
Sending the cutoff to zero and solving the equation of motion
generates a singular solution with infinite self-energy.
This is analogous to the standard way of computing the cosmological constant.
Instead ADM find that, using the general relativistic equation
of motion at finite $\epsilon$, the energy is proportional to $\sqrt\epsilon$  \cite{Arnowitt:1960zz}.
Thus, taking the limit $\epsilon\to0$,
an uncharged particle with finite ``bare'' mass
has zero physical gravitational mass and
the solution it generates is flat space.
In Becker and Reuter's calculation, the ``bare'' cosmological constant
is the analog of the bare mass.
Also in this case the equation of motion  has to be solved for finite cutoff $N$
and only then can the cutoff be removed,
leaving flat space, irrespective of the bare cosmological constant.
The fact that the ADM calculation is purely classical,
whereas the Becker-Reuter one is quantum,
does not detract very much from the analogy:
in both cases the essential point is the use of the gravitational
equation of motion before removing the regulating parameter.

The paper \cite{Becker:2020mjl} dealt with a single scalar field,
while \cite{Becker:2021pwo} dealt with quantum fluctuations of the metric itself,
where some unpleasant technical features appeared.
In this paper we consider the more general case of scalar, fermion, vector 
and symmetric tensor quantum field theories 
(corresponding, on flat space, to spin zero, one half, one and two)
on a four-dimensional sphere. 
The fields are first taken in isolation and then in arbitraty combinations,
so that the results can be applied to semi-realistic models
consisting of various combinations of free fields.
For more physical models one has to include the effects of interactions.
In the scalar case, this has been discussed  in \cite{Ferrero:2024yvw},
where it was found that putting the classical metric on shell
(i.e. solving the semiclassical Einstein equations (\ref{eineq}))
actually removes the divergences in the mass and quartic self-couplings.
This adds to the similarities with the ADM calculation,
where it was found that if the particle also has a charge,
its electromagnetic self-energy is made finite by the interaction with gravity.
\smallskip

This paper is structured as follows. In Section \ref{sec:2} we set the stage, we review the computation of the one loop effective action of fields in Euclidean de Sitter and we introduce the use of the $N$-cutoffs. Sections \ref{sec:3}-\ref{sec:6} are devoted to the treatment of scalars, fermions, Maxwell fields and gravitons, respectively. New is the inclusion of the fermions and the Maxwell fields. In Section \ref{sec:7} we consider all the kind of fields, we study different combinations and how they contribute to the effective curvature. Section \ref{sec:8} contains a discussion about a recently published paper of Branchina et al. \cite{Branchina:2024xzh}, where a similar computation is done with a different measure. Finally, we conclude in Section \ref{sec:9}.

\section{Quantum fields on Euclidean De Sitter space}\label{sec:2}

In this section we review the general logic of the calculation of \cite{Becker:2020mjl},
following the treatment in \cite{Ferrero:2024yvw}.
We assume that we are in a semiclassical regime where the metric $g_{\mu\nu}$ 
can be treated as a classical field with the (Euclidean) Hilbert action (and cosmological term)
\be
S_H(g)=\frac{1}{16\pi G}\int d^4x\sqrt{g}\left[2\Lambda-R\right]\ ,
\ee
interacting with a quantum field $\phi$ with action $S_m(\phi;g)$.
The one loop effective action (EA), that includes the effect of the backreaction
of the quantum field, is given by the somewhat abstract formula
\be
\Gamma(g,\phi)=S_H(g)+S_m(g,\phi)+\frac12\Tr\log(\Delta/\mu^2)
\label{effac}
\ee
where 
\be
\Delta=-\nabla^2+E\ ,
\label{oper}
\ee
is a Laplace-type operator, with $E$ given by the mass squared of the field,
and possibly other terms proportional to the curvature.
The mass $\mu$ a physically unimportant reference scale.
Variation of the EA with respect to the metric yields the
semiclassical Einstein equations
\be
R_{\mu\nu}-\frac12 R g_{\mu\nu}+\Lambda g_{\mu\nu}
=8\pi G \langle T_{\mu\nu}\rangle
\label{eeq}
\ee
where the l.h.s. comes from the classical Hilbert action (with a bare cosmological term)
and the r.h.s is the VEV of the EM tensor of the quantum field.

For our purposes it will be enough to study the EA on a Euclidean De Sitter space,
i.e. a sphere $S^4$.
Then, the only metric degree of freedom is the radius $r$, or equivalently
the (constant) scalar curvature $R=12/r^2$.
The equations (\ref{eeq}) reduce to the single trace equation
\be
-R+\Lambda=8\pi G\langle T^\mu_\mu\rangle\ .
\ee
We can also obtain this equation directly by deriving the EA with respect to $R$.
For example, at classical level, recalling that the volume of the four-sphere is
\be
V_4=\frac{384\pi^2}{R^2}\ ,
\ee
the Hilbert action for a spherical metric can be written
\be
S_H(R)=\frac{48\pi\Lambda}{GR^2}-\frac{24\pi}{GR}
\ee
and deriving with respect to $R$ we obtain the classical equation
\be
R=4\Lambda\ .
\ee

Aside from eliminating all gravitational degrees of freedom
except for the overall scale,
the advantage of working on a sphere is that the spectrum of Laplacians 
is well-known and therefore it is possible to calculate the EA without resorting
to heat kernel asymptotics.

The operation $\Tr$ in the definition (\ref{effac}) of the EA
is a functional trace that on the sphere can be written explicitly as a sum
over all eigenstates of the Laplacian.
We regulate the sum by putting an upper bound $N$ on the 
principal quantum number $\ell$:
\be
\frac12\Tr_N\log(\Delta/\mu^2)=
\frac12\sum_{\ell=1}^N m_\ell\log(\lambda_\ell/\mu^2)
\ .
\ee
We can compare this procedure to the more standard one of cutting off the
sum at some cutoff $C$ with dimension of mass, such that
\be
\lambda_\ell<C^2\ .
\ee
At this stage there is no significant difference between the two procedures,
because the dimensionless and the dimensionful cutoffs are simply related by
\be
C^2=\lambda_N\ .
\label{CN}
\ee
In fact, the divergences for $N\to\infty$ match those for $C\to\infty$.
However, we shall see in the following that when we demand that the background metric
satisfy Einstein's equations, the two procedures lead to very different conclusions.

\bigskip
In the next sections we specify the calculation to fields with different spins: we consider  scalar fields, fermions, vector fields and the graviton.

\section{Scalar field}\label{sec:3}

For a scalar field, the eigenvalues and their multiplicities are
\be
\lambda_\ell=\frac{R}{12}\ell(\ell+3)+E\ ,\qquad
m_\ell= \frac{1}{6}(\ell+1)(\ell+2)(2\ell+3)
\ee
where $E$ is the squared mass plus possibly a nonminimal term proportional
to $R$, and $\ell=0,1\dots$.
In the massless, minimally coupled case, the EA is
\be
\Gamma_N(R)=\frac{48\pi\Lambda}{GR^2}
-\frac{24\pi}{GR}
+\frac12\eff_S(N) \log\left(\frac{R}{12\mu^2}\right)
+\frac12\ti_S(N,0)\ ,
\label{gammaN}
\ee
where
\be
\eff_S(N)=\sum_{\ell=1}^N m_\ell= \frac{1}{12} N (N+4)(N^2+4 N + 7)
\ee
is the total number of scalar modes included in the trace and
\be
\label{ti}
\ti_S(N,z)=\sum_{\ell=1}^N m_\ell \log\left(\ell(\ell+3)+z\right)\ ,
\ee
is the quantum trace for a sphere of radius $R=12\mu^2$ and $z=12E/R$.
The functions $\eff_S$ and $\ti_S$ diverge quartically for $N\to\infty$,
In the following we shall also need the function
\be
\ess_S(N,z)\equiv\frac{\partial\ti_S(N,z)}{\partial z}
=\sum_{\ell=1}^N
m_\ell\frac{1}{\ell(\ell+3)+z}\ 
\label{ess}
\ee
that only diverges quadratically.

The equation of motion for the metric is
\footnote{This is just the trace of (\ref{eineq}).}
\be
\frac{24\pi}{GR^3}(-R+4\Lambda)
=\frac{1}{2R}\eff_S(N)\ ,
\label{eommassless}
\ee
whose solution is a sphere of curvature
\bea
R&=&\frac{24\pi}{G\eff_S(N)}\left(-1\pm\sqrt{1+\frac{G\Lambda\eff_S(N)}{3\pi}}\right)
\nonumber\\
&=& 
48\sqrt{\frac{\pi\Lambda}{G}}
\left(
\frac{1}{N^2}-\frac{4}{N^3}\right)
+\left(\frac{600\sqrt{\pi\Lambda}}{\sqrt G}-\frac{288\pi}{G}\right)\frac{1}{N^4}
+\ldots
\label{silvia}
\eea
where 
in the expansion we have selected the solution with the upper sign,
because the other gives negative curvature.
This is Becker and Reuter's main result.
We see that, opposite to standard lore,
the curvature of spacetime {\it decreases} in units of $\sqrt{\Lambda/G}$,
when more quantum modes are included in the calculation.

In the massive case, the EA for the metric becomes
\be
\Gamma_N(R)
=\frac{48\pi\Lambda}{GR^2}
-\frac{24\pi}{GR}
+\frac12\eff_S(N) \log\left(\frac{R}{12\mu^2}\right)
+\frac12\ti_S\left(N,\frac{12m^2}{R}\right)\ .
\label{gammaNxi}
\ee
The quantum trace is independent of $\phi$, so the scalar potential
does not receive any quantum corrections.
Since the sum is finite, we can take the derivative under the sum,
so that the equation of motion for the metric can be written in the form
\be
\frac{24\pi}{GR^3}(-R+4\Lambda)
=\frac{1}{2R}\left[\eff_S(N)
-\frac{12m^2}{R}\ess_S\left(N,\frac{12m^2}{R}\right)\right]\ .
\label{cadmoS}
\ee
This equation cannot be solved analytically, but one can make an ansatz
\be
R=\frac{K_2}{N^2}+\frac{K_3}{N^3}+\frac{K_4}{N^4}+\ldots
\label{gloria}
\ee
and determine the coefficients $K_n$ iteratively.
Expanding for small masses (and assuming $\Lambda>0$)
\bea\label{eq:scalar}
K_2&=&48\sqrt{\frac{\pi\Lambda}{G}}+12m^2+O(m^4)
\\
K_3&=&-192\sqrt{\frac{\pi\Lambda}{G}}-48m^2+O(m^4)
\\
K_4&=&\left(\frac{600\sqrt{\pi\Lambda}}{\sqrt G}-\frac{288\pi}{G}\right)
+128m^2+\ldots
\eea
We refer to \cite{Ferrero:2024yvw} for more coefficients.
Note that the mass-independent terms are the same as in (\ref{silvia}).

A few comments on orders of magnitude are in order: in these calculations it seems natural
to choose $\Lambda$ of order one in Planck units.
Then the coefficient $K_2$ is of the order of the square Planck mass.
If we were to choose $\Lambda=0$, there would still be
a $N^{-2}$ suppression with a much smaller coefficient $K_2$,
of the order of the square of the scalar mass.
If $\Lambda=0$ and the scalar field is also massless,
the solution with the upper sign in (\ref{silvia}) is identically zero.
The one with the lower sign has
$K_2=0$ and $K_3=0$, and $R$ decreases like $N^{-4}$,
with a coefficient $K_4$ that is again of order one in Planck units,
but negative, so not compatible
with the assumed positive curvature.
This problem is specific to the scalar field and we shall return to it later
when we discuss other matter fields.

\section{Fermion}\label{sec:4}

We consider at first the case of a Dirac field minimally coupled to gravity. 
The effective action of the system is 
\be
\Gamma_N(g,\psi )= S_H(g) + S_m(g,\psi )-\mathrm{Tr}_N\log(\cD/\mu)
\label{eaD}
\ee
where $\cD$ is the Dirac operator. 
Its eigenvalues and multiplicities are given by \cite{Camporesi:1995fb}
\be
\lambda_\ell^\pm = \pm \sqrt{\frac{R}{12}} (\ell+2)\ ,\qquad
m_\ell=4 \binom{\ell+3}{\ell}\ ,\qquad
\ell=0,1,2\dots
\ee
The trace sum is over the eigenvalues with $\ell\leq N$. 
We follow the standard method of evaluating 
$\Tr_N\log(\cD/\mu)=\tfrac12\Tr\log(\cD^2/\mu^2)$
\cite{Dona:2012am}.

On a spherical background, the effective action (\ref{eaD}) becomes
\be
\Gamma_N(R)=\frac{48\pi \Lambda}{GR^2}-\frac{24\pi}{GR}-\frac{1}{2} \eff_D(N)\log{R/\mu^2}+C(N)\,,
\ee
where
\bea
\eff_D(N)=2\sum_{\ell=0}^N m_\ell&=&
\frac12(N+1)(N+2)(N+3)(N+4)
\\
&=& \frac{1}{3}(N^4+10N^3+35N^2+50N+24)
\eea
is the total number of modes with $\ell<N$
(the factor two is due to the existence of eigenvalues with both signs).
The coefficient of the $N^4$ term is four times that of $\eff_S$,
reflecting that a Dirac spinor has four degrees of freedom.
Furthermore, extremizing the action with respect to $R$, amounts to solving
\be
\frac{24\pi}{GR^3}(-R+4\Lambda)= -\frac{1}{2R}\eff_D\,.
\ee
The solutions of this equation are
\be
R = \frac{24\pi}{G\eff_D}\left[1\pm \sqrt{1-\frac{G\Lambda\eff_D}{3\pi}} \right]
\ee
In order for the solution to be real we must have 
$\frac{G\Lambda\eff_D}{3\pi}<1$.
If $\Lambda>0$, this only happens for a finite range of values of $N$.
For negative $\Lambda$ the solutions exist for all $N$,
and the one with the upper sign goes like
\be
R= 24\sqrt{\frac{\pi|\Lambda|}{G}}\frac{1}{N^2}+O\left(\frac{1}{N^3} \right)
\ee
Remarkably, this is the same solution that we found for the scalar, aside from a factor 2.

Let us now generalize to the case of a fermion with mass $m_D$. In this case the effective action reads
\be
\Gamma_N=\frac{48\pi \Lambda}{GR^2}
-\frac{24\pi}{GR}
-\frac{1}{2}\eff_D(N)\log{\left(\frac{R}{12\mu^2} \right)}
-\frac{1}{2}\ti_D(N, z_D)\;,
\ee
where from \eqref{ti} we have introduced the function 
\be
\ti_D(N, z)=2\sum_{n=0}^{N}m_n \log{((2+n)^2+z)}
\ee
with the parameter $z=z_D$
\be
z_D= \frac{12 m_D^2}{R}\;.
\ee
The equation of motion is analogous to (\ref{cadmoS}) and is given by:
\be
\frac{24\pi}{GR^3}(-R+4\Lambda)
=\frac{1}{2R}\left(-\eff_D+z_D \mathbf{S}_D(N, z_D) \right)\;,
\ee
with the function $S_D$ defined from \eqref{ess}, namely
\be
\mathbf{S}_D (N, z)=
\frac{\partial \ti_D(N, z_D)}{\partial z}
=2\sum_{\ell=0}^{N}m_\ell\frac{1}{(2+\ell)^2+z} \;.
\ee
We can search solutions with the form 
\be
R=\frac{K_2}{N^2}+\frac{K_3}{N^3}+\frac{K_4}{N^4}+\ldots
\ee
Inserting into the equation of motion and asking for the divergences $N^6$, $N^5$ and $N^4$ to cancel, we get the equation
\be
K_2^2 - 24 K_2 m_D^2 + 576 \pi G \Lambda + 288 m_D^4 \log{\left(1 + \frac{K_2}{12 m_D^2} \right)}=0
\ee
for $K_2$, and similar ones for $K_3$ and $K_4$,
which can be solved in the limit $m_D\rightarrow0$.
Assuming $\Lambda<0$ we obtain
\bea
K_2 &=& 24\sqrt{\frac{-\pi\Lambda}{G}} + 12 m_D^2 + O(m_D^3)   
\\
K_3 &=& -120\sqrt{\frac{-\pi\Lambda}{G}} -60 m_D^2 + O(m_D^3)   
\\
K_4 &=& -480\sqrt{\frac{-\pi\Lambda}{G}} + 72\frac{\pi}{G}
+12 m_D^2\left[\frac{127}{6}+3\sqrt{-\frac{\pi}{G\Lambda}}
+\log\left(\frac{G m_D^2}{2\sqrt{-\pi G\Lambda}}\right)\right] + O(m_D^3)   
\eea

As in the case of the scalar, if we choose $\Lambda=0$
the mass of the fermion gives the dimension to the coefficient $K_2$.
However, if the fermion is massless, $K_2=0$
and the first nonzero coefficient is $K_4$, that this time is positive.
Thus the choice  $\Lambda=0$ is possible for purely
fermionic massless matter, but not for massless scalars.
\footnote{One has to take the limit $m_D\to0$ before the limit $\Lambda\to0$.}
We shall return to this point later.

\section{Maxwell field}\label{sec:5}

Now we want to consider an abelian gauge field $A_\mu$.
To fix the gauge we have to include  in the matter action a gauge fixing term for the gauge boson and the associated ghost contribution. 
The Euclidean gauge fixed action is 
\be
S_M=\int d^4x\sqrt{g}\left[\frac{1}{4}F_{\mu\nu}F^{\mu\nu}
-\frac{1}{2\alpha}(\nabla_\mu A^\mu)^2 + \nabla_\mu \bar{c} \nabla^\mu c\right]\;,
\ee
where $\alpha$ is a gauge parameter.
Commuting derivatives, the bosonic part of this action is
\be
\frac12\int d^4x\sqrt{g}A_\mu
\left[-\nabla^2 g^{\mu\nu}
+\left(1-\frac{1}{\alpha}\right)\nabla^\mu\nabla^\nu
+\frac14 Rg^{\mu\nu}\right]A_\nu\:.
\ee
The standard way to get rid of the nonminimal terms $\nabla^\mu\nabla^\nu$
in the operator is to choose the Feynman gauge $\alpha=1$.
We can avoid fixing the gauge parameter by
decomposing $A^\mu$ in its transverse and longitudinal components 
\be
A^\mu = a^\mu +\nabla^\mu \phi
\label{tl}
\ee
with 
\be
\nabla_\mu a^\mu = 0    \ .
\ee
In this way the action becomes
\be
S=\int d^4x\sqrt{g}\left[
\frac{1}{2}a_\mu\left( -\nabla^2 + \frac{R}{4}\right)a^\mu
+\frac{1}{\alpha}\phi(-\nabla^2)^2\phi 
+ \nabla_\mu \bar{c} \nabla^\mu c\right]\ .
\ee
The decomposition (\ref{tl}) is a change of variables that has
a Jacobian $J=\det'(-\nabla^2/\mu^2)$, where the prime means
that the zero mode has to be left out.

The effective action for the system is therefore given by 
\begin{align}
\Gamma_N(g,a,\phi,\bar{c},c) &= S_{EH}(g)+ S_m(g,a,\phi,\bar{c},c)
\nonumber \\  
&
+\frac{1}{2}\mathrm{Tr}_{aN}\log{\left(\left(-\nabla^2 + \frac{R}{d} \right)\frac{1}{\mu^2}\right)}
+\frac{1}{2}\mathrm{Tr}_{\phi N}\log\left((-\nabla^2)^2\frac{1}{\mu^4}\right) \nonumber\\
 &
-\mathrm{Tr}_{\bar c cN}\log\left((-\nabla^2)\frac{1}{\mu^2}\right)   
- \frac{1}{2}\mathrm{Tr}'_{JN}\log\left((-\nabla^2)\frac{1}{\mu^2}\right)\ .
\end{align}
Importantly, the contributions of the longitudinal photon and of the ghosts cancel exactly.
For the transverse field the eigenvalues and multiplicities are
\be
\lambda_\ell= \frac{\ell(\ell+3)-1}{12} R\ ,\qquad
m_\ell= \frac12\ell(\ell+3)(2\ell+3)\ .
\label{eugenie}
\ee
The Jacobian is a scalar determinant, so its calculation is based on the
eigenvalues and multiplicities given previously.

The effective action can then be expressed as
\be
\Gamma_N = \frac{48\pi\Lambda}{GR^2}-\frac{24\pi}{GR}
+\frac{1}{2}\eff_M\log{\frac{R}{12\mu^2}}
+\frac12\ti_a(N,-1)-\frac12\ti_S(N,0)\ .
\ee
Here the total number of physical degrees of freedom is
\be
\eff_M=\eff_a-\eff_S
=\frac{1}{6}N(N^3+8N^2+17N+4)    \;,
\ee
with $\eff_a$ and $\eff_S$ the total number of modes of the transverse vector
and the Jacobian, respectively.
The function $\ti_a$ is defined similarly to (\ref{ti}), but with the eigenvalues
(\ref{eugenie}).
We observe that the coefficients of the $N^4$ and $N^3$ terms
are exactly twice those of a scalar, 
reflecting the fact that a Maxwell field has two degrees of freedom.

Extremizing the effective action with respect to $R$ yields 
\be
0=\frac{24\pi}{GR^3}(R-4\Lambda)+\frac{1}{2R}\eff_M\ .
\ee
This is exactly the same equation of the scalar case, except for
the replacement of $\eff_S$ by $\eff_M$.
Again we have to choose the solution with the positive sign,
and assume $\Lambda>0$, leading to
\bea
R
&\approx& 
16\sqrt{\frac{3\pi\Lambda}{G}}
\left(
\frac{1}{N^2}-\frac{4}{N^3}\right)
+\left(\frac{-232\sqrt{3\pi\Lambda}}{\sqrt G}+\frac{96\pi}{G}\right)\frac{1}{N^4}
+\ldots
\label{silviaM}
\eea
We emphasize that the result is fully independent of the gauge parameter $\alpha$.
We interpret this as being due to the fact that the background Maxwell field
$A_\mu=0$ is a solution of the equations of motion.
Moreover, we  observe that the coefficient $K_4$ is positive,
analogously to  the case of Dirac fields.

\section{Graviton}\label{sec:6}

Next we consider the case of metric fluctuations. Also in this case we need to include a gauge fixing term and the corresponding ghost action.
Furthermore, we choose to use the exponential parametrization of the metric
\be
g_{\mu\nu}=\bar g_{\mu\rho}(e^h)^\rho{}_\nu\;,
\ee
where $h$ in the exponent is a mixed tensor $h^\rho{}_\nu$ such that 
$h_{\mu\nu}=\bar g_{\mu\rho}h^\rho{}_\nu$ is symmetric.
For the graviton field we use the York decomposition
\be
h_{\mu\nu}=h^{TT}_{\mu\nu}
+\bnabla_\mu\xi_\nu+\bnabla_\nu\xi_\mu
+\bnabla_\mu\bnabla_\nu\sigma-\frac14\bar g_{\mu\nu}\bnabla^2\sigma
+\frac14 \bar g_{\mu\nu}h\ ,
\ee
where $h^{TT}_{\mu\nu}$ is transverse and traceless,
the vector $\xi$ is transverse and all covariant derivatives
are calculated with the background metric.
It is very convenient to choose the ``unimodular'' gauge $h=0$,
that removes one of the three gauge degrees of freedom.
The remaining invariance under volume-preserving diffeomorphisms
can be fixed by the condition $F_\mu=0$, where
\be
F_\mu=
\left(\delta^\nu_\mu-\bnabla_\mu\frac{1}{\bnabla^2}\bnabla^\nu\right)
\bnabla_\rho h^\rho{}_\nu
=\left(\bnabla^2+\frac{R}{4}\right)\xi_\mu\ .
\label{gfvpd}
\ee
As discussed in \cite{book} (Section 5.4.6),
this leads to a very simple form for the (off-shell) effective action
\be
\Gamma(\bar g)=S(\bar g)
+\frac12\Tr\log\left(\Delta_{(2)}/\mu^2\right)
-\frac12\Tr\log\left(\Delta_{(1)}/\mu^2\right)\ ,
\ee
where
\be
\Delta^{(2)}=-\bnabla^2+\frac16\bR
\ee
acts on $h^{TT}_{\mu\nu}$ and
\be
\Delta^{(1)}=-\bnabla^2-\frac14\bR
\ee
acts on $\xi_\mu$.

The spectra of these operators are 
\be
\lambda_\ell^{(2)}=\frac{\ell(\ell+3)-2}{12}R\ ,\qquad
m_\ell^{(2)}=\frac56(\ell+4)(\ell-1)(2\ell+3)
\ee
with $\ell=2,3\dots$
and
\be
\lambda_\ell^{(1)}=\frac{\ell(\ell+3)-6}{12}R
\ ,\qquad
m_\ell^{(1)}=\frac12\ell(\ell+3)(2\ell+3)
\ee
with $\ell=2,3\dots$

Altogether, the effective action  can be expressed in terms of the $\ti$-functions and the number of degrees of freedom
\be
\Gamma_N = \frac{48\pi\Lambda}{GR^2}-\frac{24\pi}{GR}
+\frac{1}{2}(\eff^{(2)}-\eff^{(1)})\log{\frac{R}{12\mu^2}}
+\frac12\ti^{(2)}(N,-1)-\frac12\ti^{(1)}(N,-6)\ ,
\ee
where $\eff_G=\eff^{(2)}-\eff^{(1)}$,
\be
\eff^{(2)}(N)
=\sum_{n=2}^N m^{(2)}(n)
=\frac{5}{12}N(N-1)(N+4)(N+5)
\ee
\be
\eff^{(1)}(N)
=\sum_{n=2}^N m^{(1)}(n)
=\frac14N(N+1)(N+3)(N+4)\ ,
\ee
and similar definitions for the $\ti$-functions in \eqref{ti}.
Since the arguments of these functions are constant,
they do not affect the equations of motion and
we do not need to give them in detailed form.

The total $\eff$-function is
$$
\eff_G=\eff^{(2)}-\eff^{(1)}=\frac16(N^4-8N^3-N^2-68N)
$$
and we observe that, like in the Maxwell case,
the first two terms in the large-$N$ expansion of the function $\eff_G$ 
are the same of a free scalar,
in accordance with the count of degrees of freedom.
Thus in pure gravity, assuming $\Lambda>0$ 
\bea
R&=& 
24\sqrt{\frac{2\pi\Lambda}{G}}
\left(
\frac{1}{N^2}-\frac{4}{N^3}\right)
+\left(\frac{588\sqrt{2\pi\Lambda}}{\sqrt G}-\frac{144\pi}{G}\right)\frac{1}{N^4}
+\ldots
\label{silviaG}
\eea

Let us comment on the differences between this treatment
and the one in \cite{Becker:2021pwo}, where the linear
split of the metric was used
\be
g_{\mu\nu}=\bar g_{\mu\nu}+h_{\mu\nu}\ .
\ee
With this split, all the components of the metric in the York decomposition
are sensitive to the cosmological constant term.
In particular, the transverse traceless component is governed by the operator
\footnote{Note that this operator coincides with $\Delta^{(2)}$ on shell.}
\be
-\bnabla^2+\frac23\bR-2\Lambda\ .
\ee
The cosmological term has the same effect as a mass
and appears in the effective action through the functions $\ti(N,-24\Lambda/R)$.
This greatly complicates the analysis, because it introduces
a new dependence of the effective action on curvature.
What is more, if $\Lambda>0$, as one would naively expect,
the cosmological term has the same effect as a negative squared mass,
with the result that the poles of the function $\ti(N,-24\Lambda/R)$
appear at positive rather than negative $R$.
There are two ways of avoiding this problem.
One is to choose $\Lambda<0$, in which case the poles
occur for negative $R$, as in the case of a normal massive field.
Another way, that we have followed, 
is to use the exponential parametrization of the metric (see also \cite{Percacci:2015wwa,Ohta:2016npm} for a discussion on the parametrization),
in which case the cosmological constant only affects the propagation 
of the trace mode $h$, and only off shell.
This effect we have eliminated by the gauge choice $h=0$.
We note that if we put a gauge parameter $\alpha$ in
the gauge fixing term $\int F_\mu F^\mu$,
the result turns out to be independent of $\alpha$ \cite{book}.
This is because the sphere is ``almost on shell'',
in the sense that its metric satisfies nine out of ten Einstein's equations.
Only the overall scale factor is in general off shell.
On the other hand, if we were to repeat this calculation using the linear splitting
of the metric and another gauge,
the result would depend on the gauge parameters,
which is to be expected insofar as our background is not fully on shell.

\bigskip
Having analyzed separately all the  contributions from the different quantum fields, we will consider them together, in order to understand their interplay and study physically realistic  systems.

\section{All together}\label{sec:7}

When there is a graviton,
$N_S$ scalars with mass $m_S$, $N_D$ Dirac fields with mass $m_D$ 
and $N_M$ massless gauge fields,
the effective action reads
\begin{align}    
\Gamma_N &= \frac{48\pi \Lambda}{G R^2}-\frac{24\pi}{GR}+ 
\nonumber \\
&+\frac{1}{2}(\eff_G(N)
+ N_S \eff_S (N)
- N_D\eff_D(N) 
+ N_M\eff_M(N) 
)\log{\left(\frac{R}{12\mu^2}\right)}
\nonumber\\
&+ \frac{1}{2}(
N_S \ti_S (N, z_S) 
- N_D \ti_D (N, z_D) )
\end{align}
where $z_S=12m_S^2/R$, $z_D=12m_D^2/R$ and we omit constants.

We simplify the discussion by assuming that at fundamental level all fields are massless,
which could be the case if all masses are generated dynamically
or by a Higgs mechanism.
Then, all the $\ti$-terms in the effective action can be ignored and
the equation of motion for $R$ is very simple
\be
\frac{24\pi}{GR^3}(-R+4\Lambda)
=\frac{1}{2R}\eff_{TOT}(N)
\ee
with 
\bea
\eff_{TOT}&=&\eff_G+N_S\eff_S-N_D\eff_D+N_M\eff_N
\nonumber\\
&=&\frac{1}{12}\Delta N^4
+\frac23 \Delta' N^3
+\frac{1}{12}\Delta'' N^2
+\frac13\Delta''' N
-8 N_D\ ,
\eea
where
\bea
\Delta&=&N_S- 4 N_D + 2 N_M +2
\nonumber\\
\Delta'&=&  - N_S+ 5 N_D - 2 N_M-2
\nonumber\\
\Delta''&=& 23 N_S- 140 N_D + 34 N_M -2
\nonumber\\
\Delta'''&=&7 N_S- 50 N_D + 2 N_M-34 
\eea
In particular note that $\Delta$ is the number of bosonic minus fermionic degrees of freedom.
The solution for the Ricci scalar is given by
\bea
R^\pm
&=&\frac{24\pi}{G\eff_{TOT}(N)}\left(-1\pm\sqrt{1+\frac{G\Lambda\eff_{TOT}(N)}{3\pi}}\right)
\nonumber\\
&=&
\frac{K_2^\pm}{N^2}+\frac{K_3^\pm}{N^3}+\frac{K_4^\pm}{N^4}+O(N^{-5})\;.
\label{silviatot}
\eea
Let us now comment on the possible cases.
In previous sections we have always assumed that $\Lambda$ is positive,
for bosonic fields, or negative, for fermionic ones.
Now that both types of fields are present we do not make a priori
any such assumption.
Then we find in general for the coefficients
\bea
K_2^\pm&=& \pm48\sqrt\frac{\pi}{G}\frac{\Lambda}{\sqrt\Lambda}\frac{1}{\sqrt\Delta}
\nonumber\\
K_3^\pm&=& \pm192\sqrt\frac{\pi\Lambda}{G}\frac{\Delta'\sqrt\Delta}{\Delta^2}
\nonumber\\
K_4^\pm&=& \pm24\sqrt\frac{\pi}{G}\frac{\Lambda}{\sqrt\Lambda}\frac{48\Delta^{\prime 2}-\Delta\Delta''}{\Delta^{5/2}}
-\frac{288\pi}{G\Delta}\ .
\eea

Reality of the solution requires that the product $\Delta\Lambda$ be positive.
Thus, if bosons dominate we must choose $\Lambda>0$
and then in order to have positive curvature we have to
pick the solution $R^+$.
If fermions dominate we must choose $\Lambda<0$
and the solution $R^-$.
This is in agreement with previous assumptions.
In any case, one can achieve positive curvature,
independently of the sign of the cosmological constant in the action.

\section{Questions about the measure}\label{sec:8}

Recently there appeared the paper \cite{Branchina:2024xzh},
that also makes use of a dimensionless cutoff $N$,
but in a more traditional field-theoretic context.
In order to understand the differences, let us first briefly describe
their approach in the case of a scalar field.

The first difference is the functional measure.
Whereas in our approach the functional integration measure for the scalar field
has been implicitly assumed to be the Fujikawa measure \cite{Fujikawa:1983im,Toms:1986sh,Anselmi:1991wb}
\be
\Pi_x d\phi(x)(\det g(x))^{1/4}\mu
\ee
as carefully discussed in \cite{Becker:2020mjl},
they use the measure \cite{Fradkin:1973wke}
\footnote{Equivalent results would be obtained by
using the measure $\Pi_x d\phi(x)(\det g(x))^{1/8}$ \cite{Donoghue:2020hoh}.}
\be
\Pi_x d\phi(x)(\det g(x))^{1/4}(g^{00}(x))^{1/2}
\ee
with the result that the effective action (\ref{effac}) is replaced by
\be
\Gamma'(g,\phi)=S_H(g)+S_m(g,\phi)+\frac12\Tr\log\left(\frac{12}{R}\Delta\right)\ .
\label{effacB}
\ee
In our approach the measure must involve an arbitrary dimensionful factor $\mu$,
that also appears in the trace log formula,
whereas in \cite{Branchina:2024xzh} the measure and the operator appearing
in the trace log are automatically dimensionless,
with the result that no external scale $\mu$ is needed.
We refer the reader to \cite{Bonanno:2025xdg} for a discussion about the physical implications of the choice of measure, particularly in relation to diffeomorphism invariance.

Regardless of those considerations, if we compare the two formulas for the effective action we find that
\be
\Gamma'(g,\phi)=\Gamma(g,\phi)-\frac12\log\left(\frac{R}{12\mu^2}\Delta\right)
\eff_S(N)
\ee
and we observe that the second term exactly removes the
third term in the r.h.s. of (\ref{gammaN}).
Thus, for a massless scalar field, the equation of motion
does not receive any correction at one loop:
the solution is $R=4\Lambda$ independently of $N$.

This result holds for any massless field.
For a spinor $\psi$, a Maxwell field $A_\mu$ (including the ghosts $c$ and $\bar c$)
and a graviton $h_{\mu\nu}$ (including ghosts $\bar C^\mu$ and $C_\nu$)
the measures we implicitly used are the Fujikawa measures (all in four dimensions)
\bea
&&\Pi_x 
\left(dA_\mu(x)(\det g(x))^{1/8}\right)
\left(d\bar c(x)(\det g(x))^{1/4}\right)
\left(dc(x)(\det g(x))^{1/4}\right)
\nonumber\\
&&\Pi_x 
\left(d\bar\psi(x)(\det g(x))^{1/4}d\psi(x)(\det g(x))^{1/4}\right)
\nonumber\\
&&
\Pi_x \left(dg_{\mu\nu}(x)
d\bar C^\mu(x)(\det g(x))^{3/8}
dC_\nu(x)(\det g(x))^{1/8}\right)
\eea
whereas the measures advocated in \cite{Branchina:2024xzh} are
\footnote{Since reference \cite{Branchina:2024xzh} does not treat vector fields,
for this case we use here the measure given in \cite{Unz:1985wq}.}
\bea
&&\Pi_x 
\left(dA_\mu(x)d\bar c(x)dc(x)g^{00}(\det g(x))^{1/4}\right)
\nonumber\\
&&\Pi_x 
\left(d\bar\psi(x)d\psi(x)(\det g(x))^{3/8}\right)
\nonumber\\
&&
\Pi_x \left(dg_{\mu\nu}(x)d\bar C^\mu(x)dC_\nu(x)g^{00}(\det g(x))^{-1}\right)
\eea
As in the case of the scalar, the effect of the additional powers of the determinant
is to make the kinetic operator appearing in the $\Tr\log$'s dimensionless
and hence to remove from the effective action the terms containing
the function $\eff$, so that the equations of motion do not get any
quantum correction.

On the other hand, for massive fields
(with the bare cosmological constant playing
the role of mass term for the gravitational field) the effective action $\Gamma'$
receives contributions proportional to $\log N^2 m^4/R^2$ and $(N^2+\log N^2)m^2/R$,
that can be interpreted as quantum corrections of $\Lambda/G$ and $1/G$.
(This is in sharp contrast to our measure, with which there are no terms
proportional just to $1/R$ or $1/R^2$ in the one loop contribution.)
In the presence of scalars, fermions, and the graviton,
the effective cosmological constant and Newton constant are given by
\cite{Branchina:2024xzh}
\bea
\frac{\Lambda_{eff}}{G_{eff}}&=&
\frac{\Lambda_B}{G_B}
-\frac{N_S m_S^4}{8\pi}\log N^2
+\frac{N_D m_D^4}{4\pi}\log N^2
-\frac{3\Lambda_B^2}{\pi}\log N^2
\\
\frac{1}{G_{eff}}&=&
\frac{1}{G_B}
-N_S \frac{m_S^2}{24\pi}(N^2+2\log N^2)
+N_D\frac{m_D^2}{12\pi}(N^2-\log N^2)
+\frac{\Lambda_B}{2\pi}(N^2-8\log N^2)\ .
\eea
These formulae assume that all the particles of the same type have the
same mass. If not, the main contribution to each term comes from the heaviest
particles of each type.

In the normal QFT logic, the $N$-dependence, which gives rise
to infinitites when $N\to\infty$, has to be compensated by
the $N$-dependence of the bare parameters $\Lambda_B$ and $G_B$.
Then, $\Lambda_{eff}$ and $G_{eff}$ can be interpreted as
renormalized couplings, and they must be independent of $N$.

If the masses of the particles are negligible with respect
to the bare Planck mass, i.e. $m_S^2 G\ll1$ and $m_D^2 G\ll1$,
the dominant contribution is the gravitational one, and it gives 
\bea
\Lambda_{eff}&\approx&
12\Lambda_B
\frac{\log N}{N^2}
\\
G_{eff}&\approx&
\frac{2\pi}{\Lambda_B N^2}
\eea
One can choose $\Lambda_B$ such that $G_{eff}$ and $\Lambda_{eff}$ are 
both positive.

\section{Discussion}\label{sec:9}

The treatment of the cosmological constant we described in Sections \ref{sec:2}--\ref{sec:7} differs from the standard one in two main ways: 
the most evident one is that the ultraviolet cutoff is the dimensionless number $N$, 
rather than the usual $C$ with dimensions of mass.
This could be just a minor difference, because one can
always convert $N$ to $C$, 
as in (\ref{CN}).
It becomes more significant when combined with the second difference,
which is the use of the semiclassical Einstein equation  (\ref{eineq})
{\it before} taking the limit $N\to\infty$.
This is equivalent to using in the semiclassical Einstein equations
 the EM tensor regulated with the $N$-cutoff,
rather than a renormalized EM tensor.
This implies a different physical interpretation of the cutoff.
One normally does not think of the ultraviolet regulator as having physical meaning:
it is merely an artificial mathematical device used in intermediate steps
of the calculations, ultimately leading to renormalized quantities.
It is only the latter that are given physical meaning,
and in particular it is the renormalized EM tensor
that is used in the semiclassical Einstein equations (\ref{eineq}).
Our use of the $N$-regulated EM tensor
looks weird, if not outright wrong, from this point of view.
As already emphasized in \cite{Becker:2020mjl},
in the approach we follow here the cutoff must be regarded as having a physical meaning:
it is related to the total number of degrees of freedom in the universe.
The systems with finite $N$ are described in \cite{Becker:2020mjl}
as ``gravity-coupled approximants'',
systems that are physically realizable and can give a good description
of the real universe.

It is noteworthy that the quantum corrections do not contain any term
of the same form as the ones that are present in the classical Hilbert
action, so that there is no renormalization of the cosmological and Newton constant.
The quantum corrections are nonlocal and can be interpreted as the
terms that generate the trace anomaly in the field equations.
Another consequence is that the quantum--corrected field equations
are very different from the classical ones:
actually the quantum effects are so strong that one can have
a positive curvature even if the ``bare'' cosmological constant is negative.
This fact could have interesting applications, that we shall not enter into.

This approach is close in spirit to the old idea
that gravity may act as a universal regulator \cite{DeWitt:1964yh, Isham:1970aw, Thiemann:1997rt}.
In the case of the self-energy of a charged particle,
this has been discussed in \cite{Arnowitt:1960zz}, 
see also the recent discussion in \cite{Woodard:2024zds}.
It has indeed been shown in \cite{Ferrero:2024yvw}
that the use of the $N$-cutoff, in conjunction with Einstein's equations,
leads to finiteness of the quantum corrections to the
mass and quartic self-interaction of a scalar theory.
The full implications of this result, and the generalization
to other interacting theories, is left for future work.

This work can be extended in several directions.  First of all, the computation in a Lorentzian setting would be compelling in order to address the cosmological constant problem in a cosmological spacetime. Here,  some complications arise, due to the more involved eigenvalues and eigenstates of the Laplace operator. These should, however, reflect some physical IR properties related to the Lorentzian signature on the one side, and  give universal results in the UV limit on the other side. Secondly, it would be interesting to study related self-consistent flow equations, by taking finite differences in $N$ as iterative differential equations. Finally, this work gives a different perspective on the role of the cutoff in QFT, especially within a gravitational setting. Having  a dimensionless cutoff at hand allows to study a regularized  system without the need  to introduce any (energy) scale, opening new avenues for the interpretation of the cutoff as a physical quantity. This could be related to a quantitative notion of information of the system, encoded in the number of degrees of freedom.

\section*{Acknowledgments}
We would like to thank A. Bonanno, V. Branchina, J. Donoghue, K. Falls, L. Freidel and M. Reuter, for useful discussions and comments.
\\RF  is supported by the Friedrich-Alexander University (FAU) Emerging Talents Initiative (ETI).
RF is grateful for the hospitality of Perimeter Institute
where part of this work was carried out.
Research at Perimeter Institute is supported in part by the Government of Canada
through the Department of Innovation, Science and Economic Development and by the
Province of Ontario through the Ministry of Colleges and Universities. This work was
supported by a grant from the Simons Foundation (grant no. 1034867, Dittrich).\\


\end{document}